# First-Principles Prediction of the Softening of the Silicon Shock Hugoniot Curve

S. X. Hu (胡素兴),[1,*] B. Militzer,[2,3] L. A. Collins,[4] K. P. Driver,[2] and J. D. Kress[4]

[1]Laboratory for Laser Energetics, University of Rochester

250 East River Road, Rochester, NY 14623-1299, USA

[2]Department of Earth and Planetary Science, University of California,

Berkeley, CA 94720, USA

[3]Department of Astronomy, University of California,

Berkeley, CA 94720, USA

[4]Theoretical Division, Los Alamos National Laboratory, Los Alamos, NM 87545, USA

*E-mail: shu@lle.rochester.edu

Shock compression of silicon (Si) under extremely high pressures (>100 Mbar) was investigated by using two first-principles methods of orbital-free molecular dynamics (OFMD) and path integral Monte Carlo (PIMC). While pressures from the two methods agree very well, PIMC predicts a second compression maximum because of 1$s$ electron ionization that is absent in OFMD calculations since Thomas–Fermi-based theories lack shell structure. The Kohn–Sham density functional theory is used to calculate the equation of state (EOS) of warm dense silicon for low-pressure loadings ($P < 100$ Mbar). Combining these first-principles EOS results, the principal shock Hugoniot curve of silicon for pressures varying from ~1 Mbar to above ~10 Gbar was derived. We find that silicon is ~20% or more softer than what was predicted by widely-used EOS models. Existing high-pressure experimental data ($P \approx 1 - 2$ Mbar) seem to indicate this softening behavior of Si, which calls for future strong-shock experiments ($P > 10$ Mbar) to benchmark our results.

PAC numbers: 52.27.Gr, 51.30.+i, 64.30.-t, 52.57.-z

## I. INTRODUCTION

Silicon, an important element widely used in the semiconductor industry [1], is closely related to many other scientific fields such as geophysics [2], planetary science [3], and astrophysics [4]. Moreover, silicon has also been used as a dopant to the ablator material for target designs in inertial confinement fusion (ICF) [5–7], as well as for mitigating laser-imprint [8,9] and two-plasmon–decay effects [10] in direct-drive ICF. Although it is crucial to many scientific fields, ranging from understanding the geophysics of the Earth's outer core [2] to its application to high-energy-density physics (HEDP) and ICF, the properties of silicon at high pressures (>10 Mbar) have not yet been thoroughly studied. In the late 1960s, explosively driven shock experiments [11] only reached the highest pressure of ~2 Mbar. Laser-shock experiments in 1990s, with relatively high pressures but still below ~10 Mbar—revealed unusual behaviors of silicon under stress. For example, laser-shocked silicon exhibited a persistent thermal nonequilibrium between electrons and ions in the shock front even at ~6 Mbar pressures [12–14], while single crystals of silicon compressed by uniaxial shocks did not show the normal hydrostatic-like compression [15]. Particularly, the reduction of lattice spacing occurs only along the shock-propagation direction. These abnormalities have called for more studies on the behavior of Si under the extreme pressures of $P > 10$ Mbar.

On the theoretical side, several classical simulations [16–19] have been devoted to study the shock-wave propagation in Si. Construction of a thermodynamically complete multi-phase equation of state, in order to accurately capture the many solid-solid and solid-liquid phase transitions exhibited by silicon, provides many challenges. Classical molecular dynamics simulations [18] indicated significant deviation in shock speed from experimental measurements for $P < 1$ Mbar. Using density functional theory (DFT), Swift *et al*. [20] investigated the equation of state (EOS) of Si through the various solid-solid phase changes and then into the liquid up to 0.5 Mbar. More recently, Strickson and Artacho [21] employed a DFT/constrained molecular dynamics approach (using a "Hugoniostat") and also obtained the Si Hugoniot up to 0.7 Mbar in good agreement with the results of Ref. [20]. Additional theoretical studies of the dynamic properties of shock-compressed Si have also been performed in the pressure regime of $P < 10$ Mbar [13, 22, 23]. Most recently, Militzer and Driver [24] have extended their path integral Monte Carlo (PIMC) method to study the EOS of hot dense Si over the $P > 100$ Mbar pressure range using novel Hartree–Fock nodes. There exists another first-principles technique for hot dense matter simulations—the orbital-free molecular-dynamics (OFMD) method [25]—based on the original idea of density functional theory [26]. The OFMD method has been extensively used to simulate a variety of hot dense materials from hydrogen/deuterium [27], helium–iron mixture [28], polystyrene [29,30] to plutonium [31]; however, it has never been tested against PIMC calculations for heavy elements.

In this paper, both the OFMD and PIMC calculations of the shock Hugoniot of silicon under extreme pressures (P > 100 Mbar) have been conducted. First we determined that the two first-principles calculations, PIMC and OFMD, agree well for hot dense material simulations. Next the PIMC/OFMD results were combined with the orbital-based Kohn–Sham DFT molecular-dynamics (KSMD) calculations to derive a global Hugoniot curve of Si in the fluid and plasma phases. Finally, we benchmark these first-principles calculations with the existing experimental data and EOS models to identify the shock behavior in Si. Our results from both KSMD + OFMD and KSMD + PIMC calculations indicate that silicon is much softer than the prediction from the extensively used *SESAME* and quotidian equation-of-state (QEOS) models. The stiff behavior predicted by these EOS models was caused by the overestimation of both pressure and internal energy. The QMD-predicted softening of silicon under extreme pressure might have implications in geophysics simulations and HEDP/ICF applications. We hope these results will facilitate future high-pressure shock experiments of silicon.

The paper is organized as follows: a brief description of the three first-principles methods is given in Sec. II. Next, the derived Hugoniot curve of Si from the combined OFMD/PIMC and KSMD calculations is presented. We illustrate the softening of Si under high pressures by comparing with the widely used *SESAME*-EOS and QEOS models, as well as the existing experimental data in Sec. III. The heat capacity of shocked Si along its principal Hugoniot is also discussed in this section. Finally, the conclusions are presented in Sec. IV.

## II.  FIRST-PRINCIPLES METHODS

First-principles methods, such as DFT-based quantum molecular dynamics (QMD) [32–35], ground-state quantum Monte Carlo (QMC) [36,37], and finite-temperature PIMC [38], were developed to calculate the properties of materials under extreme conditions. The QMD method has two different implementations: (1) the orbital-based Kohn–Sham formalism [39] with finite-temperature density functional theory [40] in conjunction with the molecular-dynamics method for ion motion (denoted here as "KSMD"); and (2) OFMD [25], which is based on the original DFT idea that the free energy of a many-electron system can be written as a function depending solely on the electron density. For most cases, the KSMD method has been proven to be an accurate and efficient way to calculate material properties under high compression at temperatures generally below the electron Fermi temperature $T_F$. It becomes impractical for high-temperature ($T > T_F$) simulations because of the rapid increase in the number of orbitals required for convergence. The OFMD method is a natural extension of the KSMD method for high-$T$ materials simulations. Using the density-matrix description of a many-body quantum system, the PIMC method is also currently applicable to simulate hot dense materials of elements up to Ar. Since the details of these first-principles methods have been documented elsewhere, only a brief description is presented here.

### A.    KSMD

The KSMD method has been successfully implemented in the Vienna *ab initio* simulation package (VASP) [41–43], in which electrons are treated quantum mechanically with a plane-wave, finite-temperature DFT description. The electrons and ions of the material are in thermodynamic equilibrium with equal temperature ($T_e = T_i$).

The interaction between valence electrons and their parent ion is represented by a projector-augmented-wave (PAW) pseudopotential with "frozen" core electrons. The electron exchange-correlation potential is described by the generalized gradient approximation (GGA) with the Perdew–Burke–Ernzerhof (PBE) functional [44]. Under the Born–Oppenheimer approximation, the self-consistent electron density is first determined for an ion configuration. Next, the classical ions are moved by the combined electronic and ionic forces using Newton's equation of motion. This molecular dynamics procedure is repeated for thousands of time steps. The thermodynamic quantities such as pressure and internal energy can be calculated directly.

In KSMD simulations, the $\Gamma$ point ($\mathbf{k} = 0$) sampling of the Brillouin zone was employed. Twenty-seven Si atoms in a cubic cell with a periodic boundary condition were used. The cubic cell size was determined from the mass density. The cell-size effect has been tested by varying the number of atoms from $N = 8$ to $N = 64$; we obtained a good convergence (within ~3%) for $N \geq 16$ simulations. The PAW potential of Si included 12 valence electrons per atom. The plane-wave cutoff energy was set to 2000 eV. In all KSMD simulations, a sufficient number of bands (varying from 400 to 5000) were included such that the electron population of the highest band was less than $10^{-5}$. The time step varied from $\delta t = 1.25$ fs to $\delta t = 0.2$ fs, respectively, for the lowest to highest compressions; good convergence was obtained for these parameters.

**B.  OFMD**

The OFMD method [25] originates directly from the Hohenberg–Kohn theorem [26], with the free energy of an electron–ion system at any ion configuration can be written as a function of the electron density. In the OFMD, the electronic density is found from a finite-temperature density functional theory [25] treatment in the Thomas-Fermi-Dirac mode with the kinetic-entropic form of Perrot [45]. The electron-ion interaction is obtained from a regularization prescription [25] with a cutoff radius selected at less than 10% of the Wigner-Seitz radius to avoid the overlap of the regularized spheres. Finally, the exchange-correlation takes a local-density Perdew-Zunger form [46].

At each time step of OFMD simulations, the electron free energy for an ionic configuration is first minimized in terms of local electron density. Then the classical ions are moved by the combined electronic and ionic forces as they were in the KSMD procedure. In the OFMD simulations of Si to determine its Hugoniot, 128 atoms in a cubic cell with a periodic boundary condition were used. The small time step, varying from $\delta t = 2.4 \times 10^{-2}$ fs to $\delta t = 4.8 \times 10^{-3}$ fs, is determined by the Si density and temperature considered. Finally, the thermodynamic quantities are averaged over the MD propagation of the system (5,000 to 10,000 steps).

### C.  PIMC

Path integral Monte Carlo (PIMC) is another first-principles method for hot dense material simulations, which uses the density matrix to describe a quantum many-body system in thermodynamic equilibrium. The density matrix has a convolution property; i.e., the density matrix at $T$ can be expressed as a convolution of density matrices at very high temperatures ($M \times T$), where the correlation effects between particles are small and a very good approximation for the density matrix exists [47–49]. The low-$T$ density matrix can be found by carrying out the multidimensional integration along the imaginary time path, which is needed to recover the full correlation effects at low temperatures. The Monte Carlo method is an effective way to perform such multidimensional integrations. The electron-exchange effect is naturally taken into account through the inclusion of path permutations because the fermionic character of electrons requires an antisymmetric density matrix. The inclusion of permutation space results in inefficient sampling for lower temperatures since the positive and negative contributions to the integration are nearly cancelled out. This fermion sign problem [50] was overcome by restricting the paths using free-particle nodes or variational density-matrix nodes [51], which has succeeded in making PIMC simulations feasible to reach the lower temperatures of $T \simeq 0.1\,T_\mathrm{F}$ overlapping with the KSMD results. Such node-restricted PIMC simulations have produced accurate EOS tables of light elements such as hydrogen/deuterium [52–55], helium [56], carbon [57], water [58], neon [59], nitrogen [60], and oxygen [61].

For heavy elements such as Si, the challenge for PIMC simulations was to incorporate the effects of atomic bound states into the nodal structure. This was recently accomplished by constructing a thermal density matrix from Hartree–Fock (HF) orbitals [24]. With such an HF nodal approximation, the PIMC simulations of Si resulted in good agreement with the KSMD calculations in the temperature regime where both methods are applicable. For EOS calculations of Si, the HF-nodal PIMC simulations using eight nuclei and 112 electrons were performed at temperatures varying from $10^6$ K to $1.3 \times 10^8$ K. For such high temperatures, the PIMC simulations with 8 atoms have already given convergent results. The Hugoniot curve of Si is derived by combining the PIMC results with KSMD calculations (using 24 atoms) of silicon EOS.

## III. RESULTS AND DISCUSSIONS

The KSMD method is employed for pressures less than ~100 Mbar, whereas both OFMD and PIMC methods are used for high-temperature points with $P > 100$ Mbar. Before comparing the high-pressure Hugoniot derived from the two calculations, two tests are performed: (1) a single Si atom in a cubic box of $L = 5$ bohr, and (2) EOS calculations for Si of different densities at $T = 16,167,663$ K. Figure 1 compares the predicted pressures and energies as a function of temperature from both PIMC and OFMD calculations for the first case. The temperature varies from $T = 5 \times 10^5$ K to $T = 4 \times 10^7$ K. The pressure comparison in Fig. 1(a) indicates that both methods give perfect agreement for all temperatures except the lowest temperature point at $T = 5 \times 10^5$ K. At this low-$T$ point, the PIMC simulations predict ~1% higher pressure than the OFMD result. Figure 1(b) shows that the predicted internal energies also compare well with each other except for the temperatures $\sim T = 10^6$ to $10^7$ K, for which the two results differ by a maximum of ~6% to 10%. It is noted that because of the different potential energy references used in these methods, a constant energy shift (+92.715 Ha/atom) has been applied to all OFMD-calculated energies. The zero energy was set to the completely ionized state, which is standard in quantum-chemistry calculations. Figure 1(c) shows the relative difference of internal energy calculated by KSMD, OFMD, and PIMC. It indicates that the OFMD calculation predicts ~10% higher energy than PIMC at $T \sim 4 \times 10^6$ K, while the OFMD energies are ~6% lower than the PIMC ones around $T = 10^7$ K. This discrepancy may originate from the fact that the PIMC method handles the 1$s$ core-electron ionization explicitly through the HF-nodal implementation while the OFMD method treats atoms without shell structures. Therefore, these two treatments of electron ionization can have slightly different features in the predicted Hugoniot curve at temperatures of $\sim T = 10^7$ K (discussed below). Overall, the two methods provide reliable predictions of pressure and energy for such a simple situation.

Results from the second test case (using many particles) are presented in Fig. 2, where the quantity of $PV$ and energy are plotted as functions of Si density for $T = 16{,}167{,}663$ K. Again, the pressures predicted by both OFMD and PIMC agree well overall within ~1%. The internal energies approach each other at the low density of ~0.1 g/cm$^3$, but their difference increases gradually at high densities [see Fig. 2(b)]. Nevertheless, the predicted internal energy by OFMD and PIMC are still within ~3% at this temperature even for the highest explored density of $\rho = 46.58$ g/cm$^3$. The Debye–Hückel model (the green dashed line in Fig. 2), gives excellent agreement with both results for low densities at this high temperature but deviates from the two first-principles calculations at higher densities, especially in pressures. These test results give us more confidence in pursuing the Hugoniot comparison.

To derive the principal Hugoniot, KSMD simulations were carried out for pressures below ~100 Mbar and OFMD/PIMC calculations for extremely high pressures (>100 Mbar). The initial state chosen was solid silicon ($\rho_0 = 2.329$ g/cm$^3$) in its diamond phase at an ambient pressure ($P_0 = 1$ bar). Its internal energy of $E_0 = -289.166$ Ha/atom is determined by referring the VASP calculation of diamond-phase Si to the all-electron DFT result of an isolated Si atom. Under shock compression, the final state denoted by ($E$, $P$, $\rho$) is calculated by the following Rankine-Hugoniot equation:

$$E - E_0 + \frac{1}{2}(P + P_0)\left(\frac{1}{\rho} - \frac{1}{\rho_0}\right) = 0. \tag{1}$$

To search for the Hugoniot point for each temperature $T$, two calculations were performed at two slightly different densities $\rho_1$ and $\rho_2$. The resulting pressures ($P_1$ and $P_2$) and energies ($E_1$ and $E_2$) are applied to evaluate the Hugoniot using the above equation. If the two density points bracket the true Hugoniot density, a bisection search with further calculations may be performed to find the final Hugoniot density point. This process was repeated for a wide range of temperatures from $T = 5000$ K (above melting) to $T = 3.2 \times 10^7$ K, using both the KSMD + OFMD and KSMD + PIMC methods. The resulting principal Hugoniot of Si is shown in Table I, in which the shock and particle velocities $\left[U_s = \sqrt{\rho(P - P_0)/(\rho\rho_0 - \rho_0^2)} \text{ and } U_p = (P - P_0)/(\rho_0 U_s)\right]$ are also given. The Hugoniot results shown in Fig. 3, in which the pressure spans more than four orders of magnitude, are plotted as a function of the shock density. The red circles represent the KSMD + OFMD results, while the purple crosses represent the KSMD + PIMC calculations. Overall, the two results agree very well with each other up to ~1 Gbar pressure. In the procedure to generate the global Hugoniot by KSMD + OFMD, the "boot-strap" technique [62], has been applied to make a smooth transition from KSMD to OFMD, around $\rho = 100$ Mbar, by determining the corresponding OFMD internal energy $E_0$ of the initial Si crystal. It is noted that the obtained $E_0$ (OFMD) = −381.678 Ha/atom through the "boot-strapping" technique (plus the constant shift of +92.715 Ha/atom noted above), agrees well (within 0.2 Ha) with the all-electron DFT-calculated value of $E_0$ (KSMD) = –289.166 Ha/atom. Figure 3 shows some small differences in the pressure range of several Gbars: the PIMC simulations predict a second maximum compression resulting from the 1$s$ core-electron ionization, whereas such a feature is absent in the OFMD calculations. Namely, OFMD predicts the ionization to occur gradually along the Hugoniot, while PIMC shows distinct increases in shock compression because of the shell structure of the silicon atom. These differences can be attributed to the internal energy differences seen in Fig. 1(c) around $T = 0.6 \times 10^7$ K – $3 \times 10^7$ K, where the 1$s$ electron ionization occurs.

To further examine the physics behind the *1s*-electron ionization induced maximum compressions, we have calculated the heat capacity $C_v$ of silicon along its principal Hugoniot. As $C_v$ is a measure of the energy change rate with respect to temperature at fixed volume, one can perform two EOS calculations for each Hugoniot density ($\rho$) point with two temperatures of T+$\Delta$T and T-$\Delta$T slightly different from the Hugoniot temperature (T). The resulting internal energy difference $\Delta$E from the two calculations can be used to compute the heat capacity ($C_v \approx \Delta E/2\Delta T$) for the Hugoniot point ($\rho$,T). The resulting $C_v$ is plotted in Fig. 4 as a function of the Hugoniot pressure for both KSMD+PIMC and KSMD+OFMD calculations. As the *1s*-electron ionization process acts like a "heat-sink" for the system, one expects the heat capacity should dramatically increase during the ionization of inner-shell electrons. This is exactly what we see in Fig. 4 that the KSMD+PIMC calculation (blue dash-dotted line) gives a second peak of $C_v$ at P $\approx$ 3 Gbar, corresponding to the second compression peak seen in Fig. 3. After the *1s*-electron ionization completes, the heat capacity approaches the ideal-gas limit (black dashed line), as fully ionized Si plasma is formed. Instead of giving double peaks, the KSMD+OFMD calculation predicts a broad single peak of $C_v$. Again, this originates from the lack of shell structure in Thomas-Fermi like theories. Whether or not the ionization-induced second maximum compression can be seen in experiments remains a challenge for future extremely high pressure measurements.

Returning to Fig. 3, the first-principles results are compared with the widely used *SESAME*-EOS [63] and the QEOS models [64]. Both models are based on the chemical picture of matter, meaning that the total free energy can be decomposed into the cold curve, the ionic excitation, and the electron thermal excitation. For example, Sesame EOS models are typically constructed (constrained) by the best available experimental data (typically limited). Specifically, for Sesame 3810 (Si) constructed in 1997, the EOS below the solid-liquid phase transition was based on experimental Hugoniot data [11, 65, 66]. For conditions above the liquid phase transition, the EOS was constructed such that the shock Hugoniot was "similar" to Germanium (Sesame 3950) up to 4.4 Mbar. The ion thermal contribution is based on a Debye model with a correction for the liquid specific heat beyond the melt temperature [67]. The correction also ensures that in the high temperature limit the proper model (ideal gas) is recovered that gives a shock Hugoniot compression ratio $\rho/\rho_0=4$. This comparison in Fig. 3 indicates that under shock compression silicon is much softer than predicted by the traditional chemical-picture understanding of materials. For a shock density of $\rho$ = 5 to 9 g/cm$^3$, the pressures predicted from our first-principles calculations are smaller than the corresponding *SESAME* and QEOS pressures by a factor of ~2 to 3. In other words, for the same shock pressure in the range of 5 to 100 Mbar, the first-principles–predicted shock density is ~20% higher than the *SESAME* and QEOS prediction. The maximum compression ($\rho/\rho_0$) changes from the model-predicted value of ~4.0 (QEOS) and ~4.6 (*SESAME*) to ~5.0 in our first-principles calculations. Perhaps in the case of SESAME this amount of difference is not surprising based on the way that SESAME 3810 was constructed. For QEOS, it appears that the EOS is constructed so to follow the Hugoniot until the ideal gas limit (four fold compression), at which point the Hugoniot climbs nearly vertically in pressures. Looking at a typical density/temperature condition along the principal Hugoniot (e.g., $\rho$=6.5 g/cm$^3$ and T=62500 K), we find the SESAME 3810 predicted pressure and internal energy (P≈11.45 Mbar and E≈34.5 eV/atom) are about ~45% higher than our first-principles calculations (P≈7.76 Mbar and E≈24.3 eV/atom). The overestimation of pressures and energies in SESAME 3810 is the cause of the stiff behavior we saw in Fig.3. Finally, the earlier experimental data [11] and other Hugoniot measurements in 1970s and 1980s in the liquid phase [65, 66] are put into the same figure for comparison, respectively represented by the open triangles, diamonds, and squares in Fig. 3. These Hugoniot data were obtained from explosively driven shock experiments. To the best of our knowledge, no published data exists for Hugoniot measurements in pressures above 10 Mbar. The opaqueness of Si for most VISAR (velocity interferometer system for any reflector) laser wavelengths [68] is one of the hurdles for accurate shock measurements in silicon. Since some of these explosively driven shock data at low-pressures (P<1 Mbar) were used to constrain the construction of SESAME 3810, the agreement there was guaranteed. The recent KSMD data by Strickson and Artacho [21] also agree with the data at these low pressures. However, for pressures between 1 and 2 Mbar our KSMD calculations are in better agreement with the available high-pressure data [11]; while the SESAME-EOS, guided by the Ge-based model, results in twice the pressure compared to the KSMD result. It was noted that at the measured highest shock density of $\rho = 4.6$ g/cm$^3$, the *SESAME* and QEOS predicted pressures are about twice higher than both our KSMD results and the experimental value of $P \simeq 1$ to 2 Mbar. It is expected that future experimental data in the 10- to 100-Mbar pressure regime will unambiguously benchmark the predicted softening of Si.

To further understand the shock-induced structure change, we turn to the pair correlation function along the predicted principal Hugoniot of Si (Fig. 5). A strong peak at the Si–Si inter-particle distance of ~2.25 bohr exists for compression at $T = 5000$ K and $\rho = 4.38$ g/cm$^3$. As the shock compression increases to $\rho = 5.50$ g/cm$^3$ and $T = 31{,}250$ K, this peak moves to a smaller ionic distance, and the amplitude decreases. Increasing the shock strength further, this peak becomes much less pronounced at $\rho = 7.32$ g/cm$^3$ and $T = 125{,}000$ K, which manifests the status of the structureless fluid phase. Finally, the first-principles predicted shock temperatures are compared with the *SESAME* and QEOS models in Fig. 6. For shock pressures exceeding ~3 Mbar, the first-principles–predicted shock temperatures are ~30% higher than the *SESAME* EOS values and much higher (~50% to 200%) than the QEOS predictions. At a fixed pressure, the predicted shock from first-principles calculations, reaching both higher densities and temperatures, can have implications in HEDP and ICF applications. This is because the shock velocity from first-principles calculations can be significantly lower than the widely used EOS models, while the FP-predicted particle velocity is higher than *SESAME* and *QEOS* models.

## IV. SUMMARY

Two first-principles methods of PIMC and OFMD have been used to calculate EOS of silicon at extremely high pressures. Combining these high-temperature calculations with the KSMD simulation of low-temperature EOS, the principal Hugoniot curve of silicon for pressures varying from ~1 Mbar to above 10 Gbar was derived. Overall, the two results agree very well with each other, although a small difference exists at pressures around several Gbar. In this high-pressure regime, the 1$s$ core-electron ionization of Si can cause a maximum compression, since the ionization process acts like a heat sink. This is evidenced by the peaked heat capacity in this pressure range of several Gbar, in which $C_v$ is larger than the ideal-gas limit of fully-ionized Si plasma. When the first-principles predicted Hugoniot of Si is compared with the widely used *SESAME* and QEOS models, we find that silicon under pressure is ~20% or more softer than what was generally believed in the chemical picture of matter. In future work, the broad range of first-principles data will be incorporated into an improved Sesame model, in similar fashion to recent new models for the lithium deuteride [62] and multiphase germanium [69] equations of state. For the same shock density, the EOS models predicted pressures about ~2 to 3× higher than our KSMD + OFMD/PIMC results. The existing experimental data at the highest pressure of $P \approx 2$ Mbar seem to point to the same softening feature as predicted. Finally, the first-principles calculations predict an ~30% or more higher shock temperature than the *SESAME* and QEOS models at pressures of 5 to 100 Mbar. The softening of Si under high pressures can affect the geophysics simulations and the shock propagation in HEDP and ICF applications. We hope these calculations will facilitate future high-pressure experiments in the 10- to 100-Mbar regime.

**ACKNOWLEDGMENT**

This material is based upon work supported by the Department of Energy National Nuclear Security Administration under Award Number DE-NA0001944, the University of Rochester, and the New York State Energy Research and Development Authority. The support of DOE does not constitute an endorsement by DOE of the views expressed in this article. This work was also supported by Scientific Campaign 10 at the Los Alamos National Laboratory, operated by Los Alamos National Security, LLC for the National Nuclear Security Administration of the U.S. Department of Energy under Contract No. DE-AC52-06NA25396. BM and KPD acknowledge support from the Department of Energy, Grant Number DE-SC0010517. We would like to acknowledge discussions with Drs. Scott Crockett, Carl Greeff, and Daniel Sheppard of Los Alamos National Laboratory and thank DS for generating the SESAME 3810 Hugoniot using Los Alamos tools.

**FIGURE CAPTIONS**

FIG. 1. (Color online) The pressure and energy of a single Si atom inside a 5-bohr cubic box, predicted by the two different methods of PIMC and OFMD, are plotted as a function of temperature.

FIG. 2. (Color line) The comparison of pressure-related quantity *PV* and internal energy as a function of Si density at $T = 16,167,663$ K. The OFMD and PIMC results are compared with the Debye–Hückel model.

FIG. 3. (Color online) The principal Hugoniot of silicon derived from first-principles calculations compared to the two different models (*SESAME*-EOS and QEOS) and the existing experimental data by Pavlovskii [11], by Gust and Royce [65], and by Goto *et al* [66]. Also, the recent KSMD results by Strickson and Artacho [21] at lower pressures are also shown. The diamond-phase solid silicon was chosen as the initial state having a density of $\rho_0 = 2.329$ g/cm$^3$.

FIG. 4. (Color online) The comparison of heat capacity C$_v$ of silicon along its principal Hugoniot predicted by the two methods of KSMD+PIMC (dash-dotted) and KSMD+OFMD (solid). The dashed line represents the ideal-gas limit of fully ionized Si.

FIG. 5. (Color online) The pair-correlation functions along the principal Hugoniot at different compressions from KSMD simulations.

FIG. 6. (Color online) Comparison of the shock temperatures along the principal Hugoniot between our first-principles calculations and the two different *SESAME*-EOS model (*SESAME* and QEOS) predictions.

TABLE I. The principal Hugoniot of silicon (diamond phase) with an initial density of $\rho_0 = 2.329$ g/cm$^3$, predicted by KSMD + OFMD calculations.

| $T$ (K) | $\rho$ (g/cm$^3$) | $P$ (Mbar) | $U_s$ (km/s) | $U_p$ (km/s) | $\rho/\rho_0$ |
|---|---|---|---|---|---|
| 5,000 | 4.378 | 1.111 | 10.10 | 4.73 | 1.88 |
| 15,625 | 4.954 | 2.183 | 13.30 | 7.05 | 2.13 |
| 31,250 | 5.497 | 3.786 | 16.80 | 9.68 | 2.36 |
| 62,500 | 6.357 | 7.428 | 22.44 | 14.22 | 2.73 |
| 125,000 | 7.316 | 15.420 | 31.17 | 21.24 | 3.14 |
| 250,000 | 8.586 | 35.072 | 45.46 | 33.13 | 3.67 |
| 500,000 | 10.493 | 96.091 | 72.82 | 56.66 | 4.51 |
| 750,000 | 11.173 | 177.015 | 97.99 | 77.56 | 4.80 |
| 1,000,000 | 11.374 | 257.486 | 117.91 | 93.76 | 4.88 |
| 2,000,000 | 11.663 | 642.429 | 185.65 | 148.58 | 5.01 |
| 3,000,000 | 11.674 | 1085.54 | 241.30 | 193.16 | 5.02 |
| 4,000,000 | 11.607 | 1556.61 | 289.16 | 231.14 | 4.98 |
| 6,000,000 | 11.341 | 2528.47 | 369.62 | 293.72 | 4.87 |
| 8,000,000 | 11.218 | 3489.69 | 434.85 | 344.57 | 4.82 |
| 10,000,000 | 11.043 | 4436.17 | 491.31 | 387.69 | 4.74 |
| 16,000,000 | 10.649 | 7175.98 | 627.98 | 490.64 | 4.57 |
| 32,000,000 | 10.122 | 14,097.20 | 886.67 | 682.66 | 4.35 |

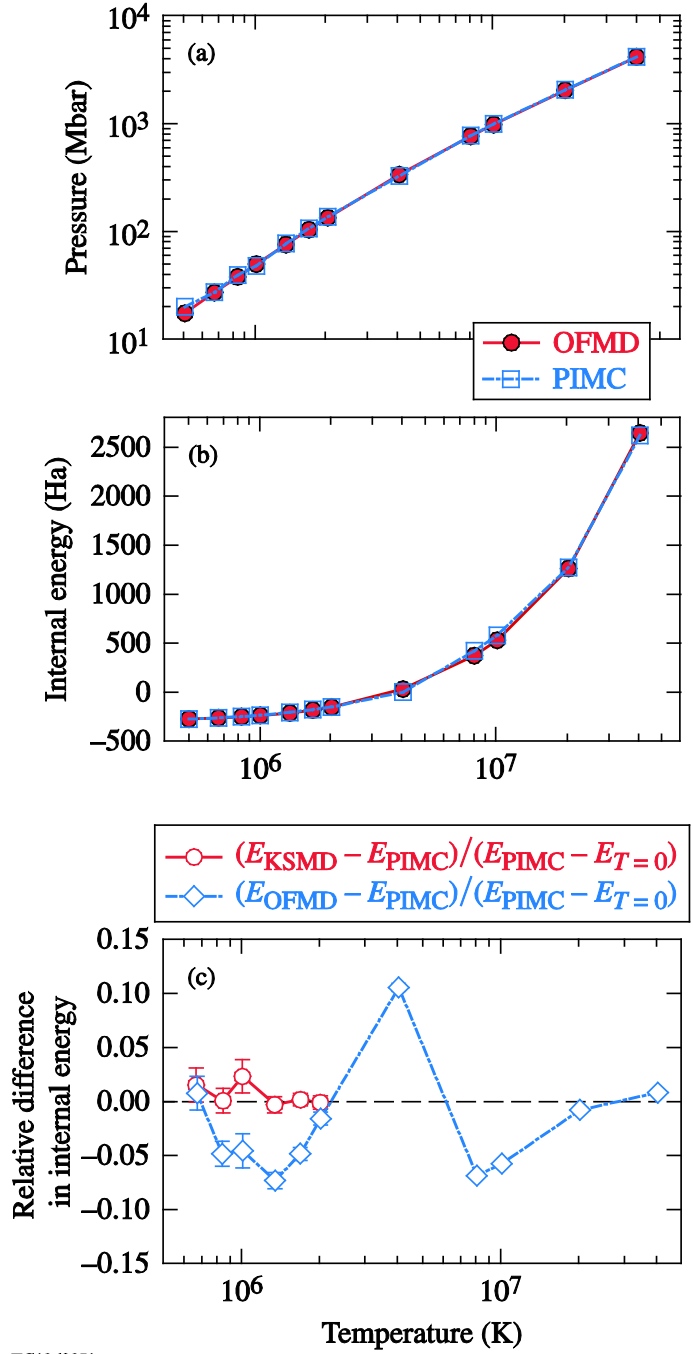

FIG. 1. (Color online) The pressure and energy of a single Si atom inside a 5-bohr cubic box, predicted by the two different methods of PIMC and OFMD, are plotted as a function of temperature.

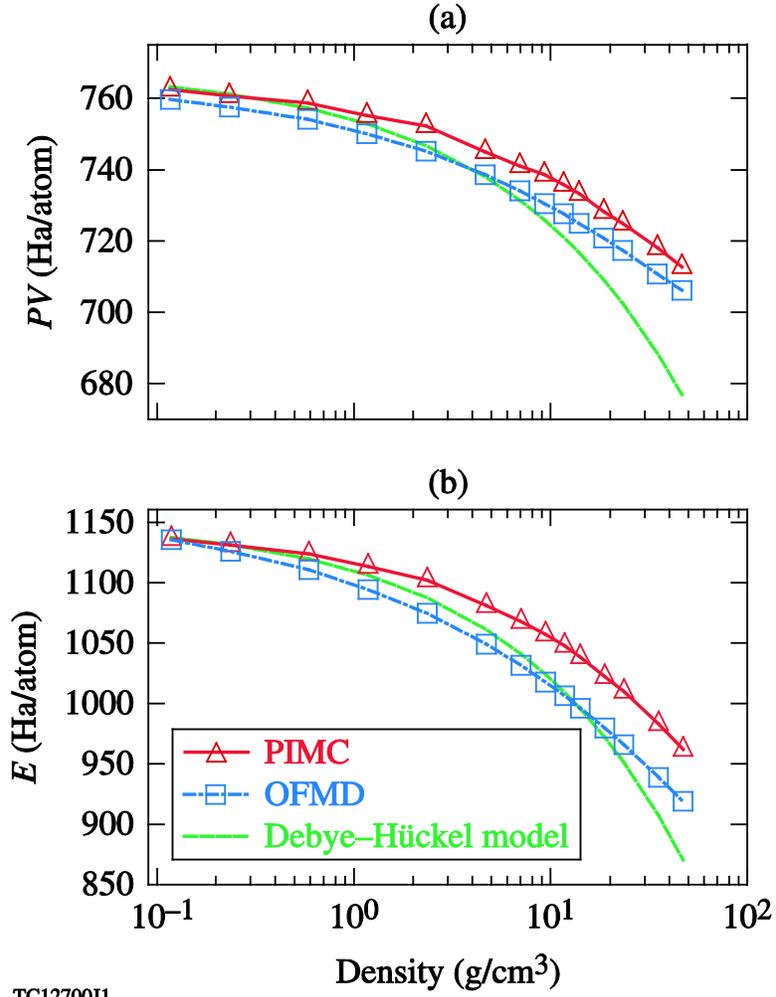

FIG. 2. (Color line) The comparison of pressure-related quantity *PV* and internal energy as a function of Si density at $T = 16{,}167{,}663$ K. The OFMD and PIMC results are compared with the Debye–Hückel model.

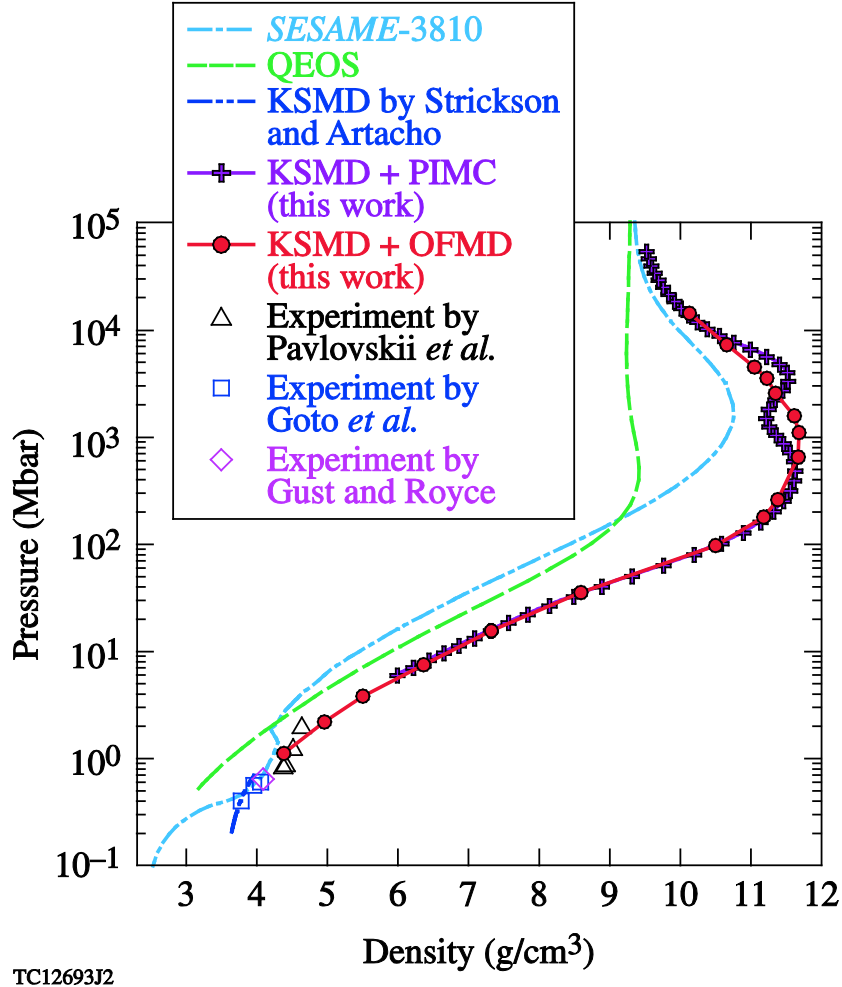

FIG. 3. (Color online) The principal Hugoniot of silicon derived from first-principles calculations compared to the two different models (*SESAME*-EOS and QEOS) and the existing experimental data by Pavlovskii [11], by Gust and Royce [65], and by Goto *et al* [66]. Also, the recent KSMD results by Strickson and Artacho [21] at lower pressures are also shown. The diamond-phase solid silicon was chosen as the initial state having a density of $\rho_0 = 2.329$ g/cm$^3$.

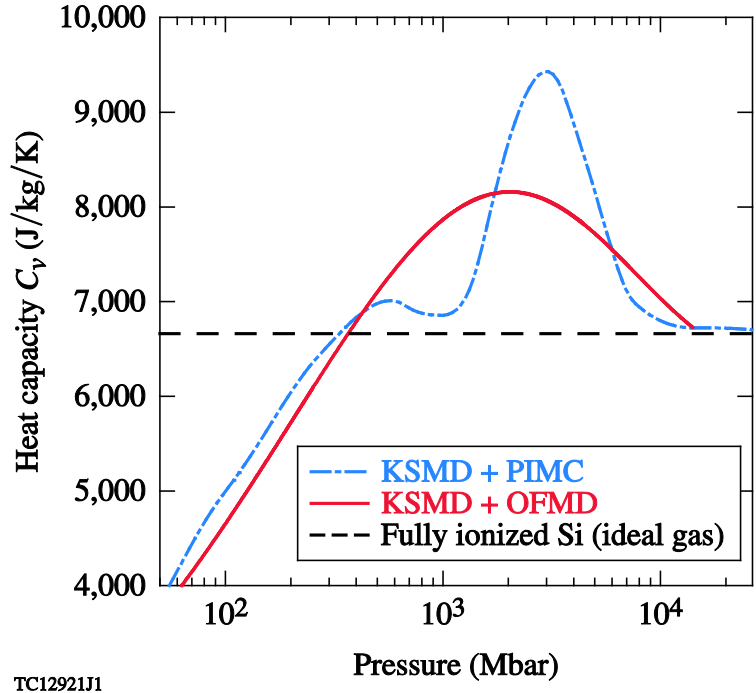

FIG. 4. (Color online) The comparison of heat capacity $C_v$ of silicon along its principal Hugoniot predicted by the two methods of KSMD+PIMC (dash-dotted) and KSMD+OFMD (solid). The dashed line represents the ideal-gas limit of fully ionized Si.

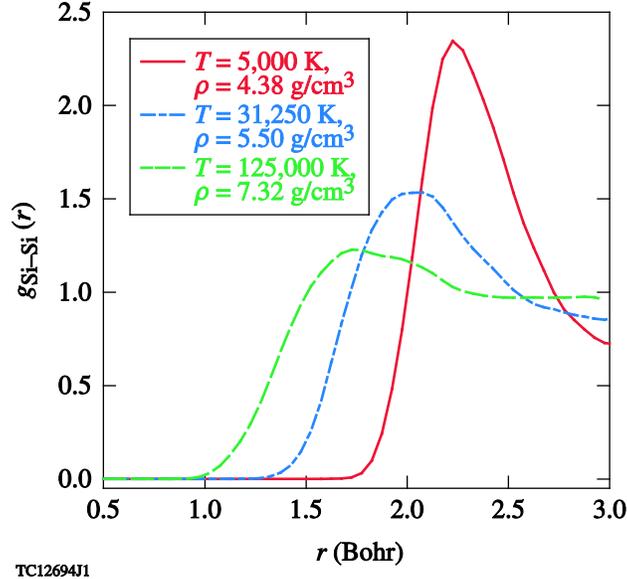

FIG. 5. (Color online) The pair-correlation functions along the principal Hugoniot at different compressions from KSMD simulations.

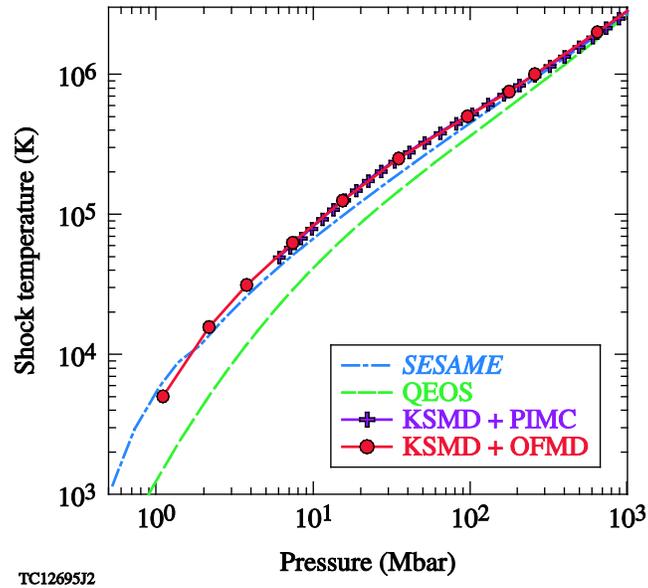

FIG. 6. (Color online) Comparison of the shock temperatures along the principal Hugoniot between our first-principles calculations and the two different *SESAME*-EOS model (*SESAME* and QEOS) predictions.